\documentclass[epsf]{mn2e}
\def\mmm{(m-M)$_0$}
\def\ebv{$E(B-V)$~}

\def\gsim{\;\lower.6ex\hbox{$\sim$}\kern-7.75pt\raise.65ex\hbox{$>$}\;}
\def\lsim{\;\lower.6ex\hbox{$\sim$}\kern-7.75pt\raise.65ex\hbox{$<$}\;}

\title[Pismis 2]{Pismis 2, a poorly studied, intermediate age open cluster\thanks{
 Based on observations collected at the European Southern Observatory, Chile}}

\author[Di Fabrizio et al.]{L. Di Fabrizio$^1$\thanks{
 Present address: Telescopio Nazionale Galileo,
 38700 Santa Cruz de La Palma, Spain,
 e-mail difabrizio@tng.iac.es}, 
A. Bragaglia$^1$, M. Tosi$^1$,  G. Marconi$^{2,3}$ \\
 $^1$ Osservatorio Astronomico di Bologna, Via Ranzani 1, I-40127 Bologna,
      Italy,
      e-mail angela, tosi @bo.astro.it \\
  $^2$ Osservatorio Astronomico Roma, Via dell'Osservatorio 5, I-00040 Monte
      Porzio, Italy, \\
  $^3$ ESO, Alonso de Cordova 3107, Vitacura, Santiago, Chile,
      e-mail gmarconi@eso.org}

\date{}

\begin{document}
\maketitle

\begin{abstract}
We present CCD $BVI$ photometry of the intermediate age open cluster Pismis 2,
covering from the brighter red giants to about 5 magnitudes below the main
sequence turn-off. The cluster is heavily reddened and is possibly affected
by a differential reddening of $\Delta E(B-V) \simeq$ 0.04--0.06.

Using the synthetic Colour - Magnitude Diagram method, we estimate in a
self-consistent way distance modulus, \mmm $\simeq$ 12.5--12.7, and age,
$\tau \simeq$ 1.1-1.2 Gyr. The cluster probably contains a significant
fraction of binary systems. The metallicity is most likely solar and the
reddening \ebv ranges between 1.26 and 1.32 depending on the cluster region.
\end{abstract}


\section{Introduction}

The r\^ole of old and intermediate age open clusters (OCs) in probing the
chemical and dynamical evolution of the Galactic disc has long been
recognized (e.g., Panagia \& Tosi 1981, Friel 1995). 
Over the last years, several
authors have collected  a considerable amount of high quality photometric
data, but to fully exploit OCs' potentialities, results have to be based on
a sample as complete (to ensure significance) and as homogeneous (to avoid
spurious effects due to the use of different techniques) as possible.

To this end, we are homogeneously analysing with great accuracy a sample of
open clusters at various galactic locations and covering a wide range in age
and metallicity, in order to significantly sample the disc properties both
in space and time. Up to now we have published results for NGC 7790, NGC
2660, NGC 2506, Berkeley 21, NGC 6253, NGC 2243, and Collinder 261, in order
of increasing age, from 0.1 to at least 7 Gyr (see Sandrelli et al. 1999,
Bragaglia et al. 2000 and references therein).
Age, distance, reddening and approximate metallicity are derived from deep
photometry combined with the synthetic Colour-Magnitude diagrams method 
(Tosi et al. 1991). Whenever possible, we
complement the photometric study with high resolution spectroscopy, to derive
accurate metal abundances (Carretta et al. 2000, Bragaglia et al. 2001).

This paper is part of our general project of observations of galactic open
clusters; we present here results on the open cluster Pismis 2
(Pismis 1959), or C0816-414, with coordinates $\alpha_{2000} = 8^h18^m3^s$,
$\delta_{2000}= -41^o 37\arcmin 25\arcsec$, l=258.83, b=--3.29.

To our knowledge, the first and only colour magnitude diagram published for
this object is by Phelps, Janes \& Montgomery (1994, PJM94, their fig. 19);
they observed in the $V$ and $I$ filters, and estimated an age of 1.1 Gyr,
based on the MAI (Morphological Age Indicator, i.e. the difference in
magnitude between the main sequence turn-off and the red clump). In an
associated paper on the general properties of the galactic old OCs, Janes \&
Phelps (1994) give for this cluster \ebv = 1.48, as deduced from the mean
colour of the red clump. No other measured values for reddening or 
metallicity could
be found in literature. Patino \& Friel (1994) presented preliminary results
on its kinematics, based on medium resolution spectroscopy, but never
published the figures; the same did Miller et al. (1995), who included
Pismis 2 in a list of open clusters for which reddening could be obtained
through selected spectral features.
The latest interstellar absorption maps (based on FIR data) by Schlegel, 
Finkbeiner \& Davis (1998) according to  the authors  cannot be trusted to 
give a reliable \ebv value, because of the very low galactic latitude of the
cluster.  However,  Dutra \& Bica (2000) compared reddenings derived from those
maps and from photometry of star clusters, both globular and open, and found
reasonable agreement also at low Galactic latitudes, except  for some objects,
among them Pismis 2, for which the FIR maps provided higher values
(in our case, \ebv = 2.27). According to them, most of the discrepancy between
the reddening values must be ascribed to dust clouds in the disc background of
the clusters.

We describe the data acquisition and reduction in Section 2, the resulting
colour-magnitude diagrams (CMDs), membership and reddening
in Section 3. The cluster parameters are derived from the synthetic
CMDs in Section 4 and a summarizing discussion is given in Section 5.

%%%%%%%%%%%%%%%%%%%%%%%%%%%%%%% Fig. 1 (MAP)
\begin{figure}
\vspace{9cm}
\includegraphics{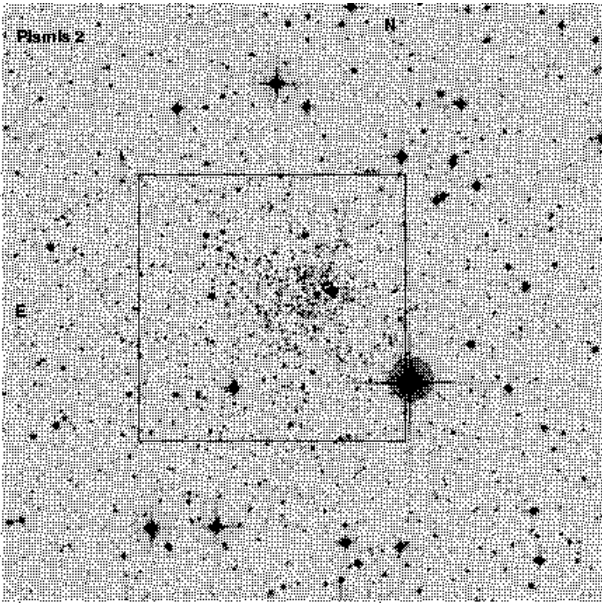}
\caption{Map of our field (the central square) of 6.4 arcmin$^2$; the
whole figure is 15 arcmin$^2$, and is a good approximation of PJM94 field of
view (14 arcmin$^2$).}
\label{fig-map}
\end{figure}

\section{Observations and data reduction}

Pismis 2 was observed at the 1.54m Danish telescope located in La Silla,
Chile, on UT March 10, 1995; we have only one field, centred on the cluster.
The instrument used was the direct camera, mounting the CCD \#28, a Tek
1024$\times$1024, with a scale of 0.377 \arcsec/pix, yielding a field of view
of 6.4 arcmin$^2$.
The observed region is indicated by the central square in Fig.~\ref{fig-map},
taken from the DSS, of 15 arcmin$^2$ (more or less the PJM94 field), and
oriented with North up and East left.

We took exposures ranging from 60 to 1800 seconds in $B$, and from 20 to 900 
seconds in $V$ and $I$, all in a 2 hours time span. Standard areas were
observed just after the cluster and later during the night. Conditions were
photometric, and seeing was $\lsim$ 1\arcsec.

Standard CCD reduction of bias subtraction, trimming, and
flatfield correction were performed.
We applied to all frames the usual procedure  for PSF study and fitting
available in DAOPHOT--II (Stetson 1992) in IRAF\footnote{
IRAF is distributed by the NOAO, which are operated by AURA, under contract
with NSF} environment.
All frames were searched independently, using the appropriate value for the
FWHM of the stellar profile and a  threshold of 3 to 4 $\sigma$ over the local
sky value. We used different thresholds (3, 3.5, and 4 $\sigma$ in $B$, $V$,
and $I$ respectively) to keep them as low as possible but avoiding false
identifications due to the spikes of the saturated stars, that happened to
be more numerous in the $V$ and $I$ images.
In all frames, we also made a second pass with a higher threshold (6 $\sigma$)
to try to recover hidden components.  

About 20 well isolated, bright stars were used in each frame 
to define the best analytical PSF model (in our case, a "penny1" without spatial
variations), which was then applied to all detected objects.
The resulting magnitude file was selected both in magnitude, to avoid saturated
stars, and in sharpness, a shape - defining parameter, to avoid cosmic rays and
false identifications of extended objects. Areas around badly saturated 
stars were blanketed, i.e. we eliminated from the output catalogues
all the objects falling within a radius of 20 pixels from the saturated stars. 

All steps were archived as an IRAF procedure, to be reproduced during the
completeness study.

All output catalogues were aligned to the one derived from the deepest $I$
image, assumed as master frame for the coordinate system, using dedicated
programs developed at the Bologna Observatory by P. Montegriffo. 

All output catalogues in each filter were "forced" on the deepest
one in that band simply applying a "zero point" shift to the instrumental
magnitude (ranging from a few thousandths to about one tenth of magnitude). 
The final magnitude catalogue in each band is the result of the
(weighted) average of all measures for each stars

We computed a correction to the PSF derived magnitudes to be on the
same system as the photometric standard stars:
aperture photometry was performed on a few isolated stars (the same used to
define the PSF) in the  reference images, i.e., in the deepest frame for each
filter. The correction to apply to the PSF magnitudes was found to be:
--0.272 in $B$, --0.232 mag in $V$, and --0.233 in $I$ (in the sense aperture
minus PSF).

\subsection{Photometric calibration}

The conversion from instrumental magnitudes to the Johnson-Cousins
standard system was obtained using a set of primary calibrators
(the areas Rubin 149, PG0918+029, PG1047+003, PG1323--086, PG1633+099, for a
total of 26 stars: Landolt 1992). The stars defined by Landolt
span a fairly wide range in colour (--0.290 $\leq B-V \leq$ +1.134,
--0.295 $\leq V-I \leq$ +1.138), 
but the cluster is heavily reddened, 
and our calibration is all in extrapolation using these stars.
Recently, Stetson (2000) has presented an extension to fainter magnitudes
of Landolt's standard stars fields. We have tried to recover in our frames at
least some of the redder stars, to extend the colour baseline of our
calibration. Only a few stars are usable, though, because of their general
faintness. We could recover 6 stars in $V$ and $I$ (PG0918-S14, and S29;
PG1047-S12, and S15;  PG1323-S18, and S27), thus extending our colour baseline
to $V-I$ = 2.263. Only 2 of these stars, none redder than our previous limit in
instrumental $(b-v)$,  were measurable in the  $B$ frames; they were not
considered in the calibration of the $B$, since their $B$ magnitudes were not
given by Stetson, while  they were included in the $V$ $vs$ $(b-v)$
calibration.

\begin{figure}
\vspace{12cm}
\includegraphics{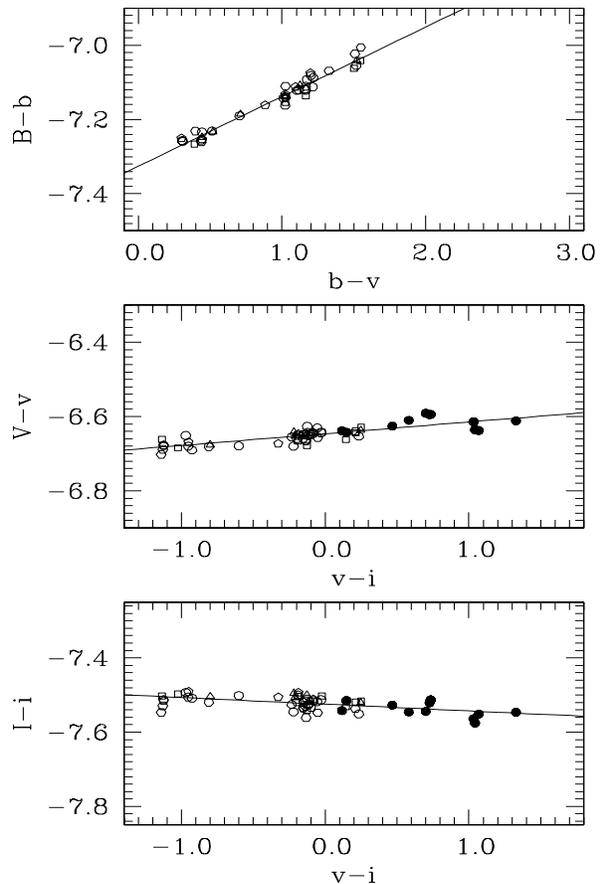}
\caption{Calibration equations, as given in the text. Open symbols: Landolt
stars; filled symbols: Stetson stars. Note that the reddest Landolt star 
has $B-V$=1.134 and $V-I$=1.138, while the 
reddest Stetson star has $V-I$=2.263 .}							    
\label{fig-cal}
\end{figure}

Standard stars fields were analysed using aperture photometry. The
calibration equations were derived using the extinction coefficients for La
Silla taken from the database maintained by the photometric group at the
Geneva Observatory Archive.
Since all the 6 nights in the same run were photometric, we derived a unique
calibration using the whole set of standards (Landolt's plus Stetson's) 
and obtained equations in the form:
$$ B = B +0.187 \cdot (b-v) -7.326 ~~~~(r.m.s.=0.014) $$
$$ V = V +0.037 \cdot (b-v) -6.693 ~~~~(r.m.s.=0.012) $$
$$ V = V +0.032 \cdot (v-i) -6.646 ~~~~(r.m.s.=0.014) $$
$$ I = I -0.018 \cdot (v-i) -7.528 ~~~~~(r.m.s.=0.016) $$
where $b,v,i$ are instrumental magnitudes, while $B,V,I$ are the
corresponding Johnson-Cousins magnitudes. Figure~\ref{fig-cal} shows three of
these equations; open symbols represent Landolt stars, and filled ones Stetson
objects.
 
We also checked that the calibration of each night taken separately does not
differ from the above relations. As already said, an extension  to colours as
red as the cluster stars was possible only for the $V$ and $I$ filters.  We
note however that the extended ("Landolt" plus "Stetson") calibration 
equations  in these bands produce magnitudes that differ from the traditional 
("Landolt") ones only by  a few thousandths of magnitude, hence the 
extrapolation we are forced to adopt for the $B$ band looks quite reasonable
(see, however, Section 5).  
We calibrated the $B$ magnitudes using the relation involving
$(b-v)$,  and the $I$ and $V$ magnitudes with the ones involving $(v-i)$
(except for the  few  tens of stars missing $i$, for which we had to use the
$(b-v)$ colour to  calibrate $V$).

%%%%%%%%%%%%%%%%%%%%%%%% Fig. 2 (eq. calibr.)

%%%%%%%%%%%%%%%%%%%%%%%%% Fig. 3 (errors)
\begin{figure}
\vspace{15cm}
\includegraphics{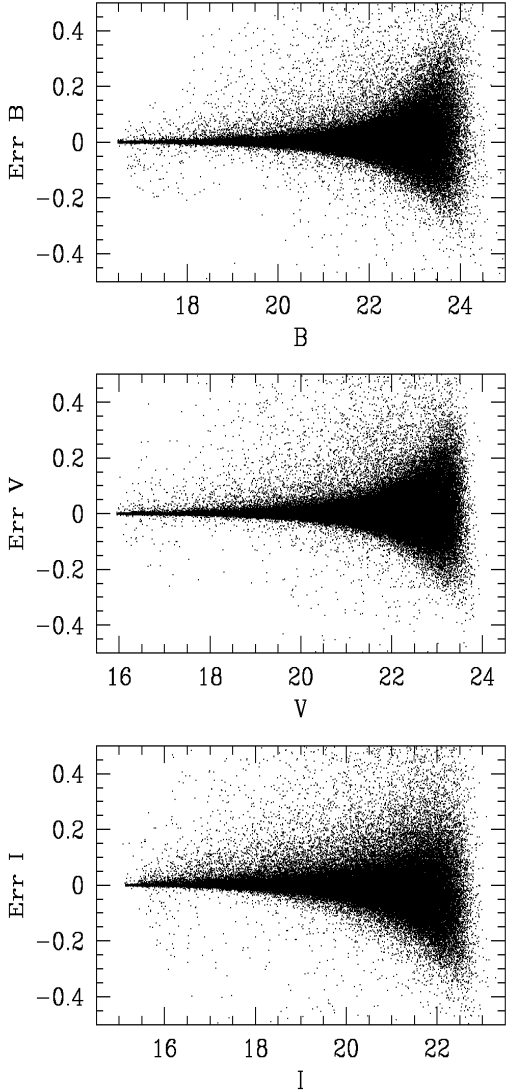}
\caption{Difference in magnitude (input minus output) of the artificial
stars in the completeness test in all the three filters.}
\label{fig-ref1}
\end{figure}
						    
%%%%%%%%%%%%%%%%%%%%%%%%% Fig. 4 (comparison with PJM94)
\begin{figure}
\vspace{10cm}
\includegraphics{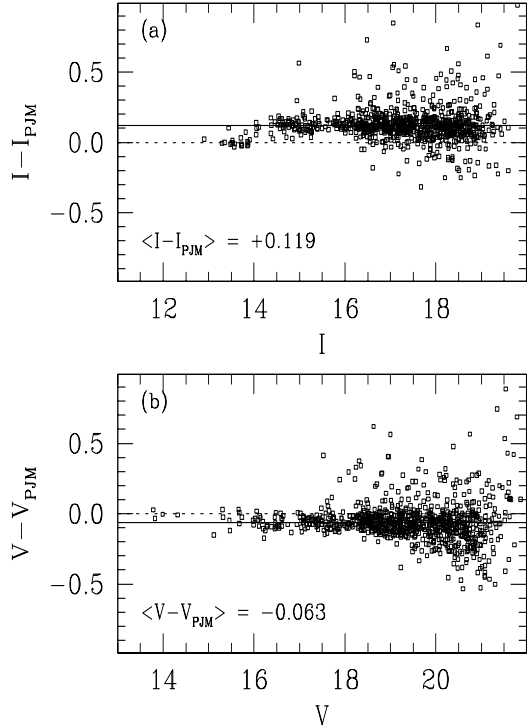}
\caption{(a) Comparison between our values for the $I$ magnitudes and those by
PJM94: we are on average fainter by 0.119 mag; (b), the same, in $V$: we
are on average brighter by 0.063 mag.}
\label{fig-conf}
\end{figure}

\subsection{Completeness analysis}

We tested the completeness of our luminosity function in the $B, V$ and $I$ 
band on the deepest images, adding artificial stars to the frames  and
exactly repeating the procedure of extraction of objects and PSF fitting
used for the original frame. The stars were added at random positions 
and selected in magnitude according to the observed
luminosity  function. We added 120 objects at a time, 
in order not to significatively alter the crowding conditions, repeating  the
process as many times as to reach a total of about 200,000 artificial stars. 
This way we approximate the condition of adding a single star each time, i.e.
of a repeated, independent experiment, since the few artificial objects 
added in a single run do not interfere with each other.
To the output catalogue of the added stars we applied the same selection criteria
in magnitude and sharpness as done for the science frames.
The completeness degree of our photometry at
each magnitude level was computed  as the ratio of the number of
recovered stars to the number of simulated ones  (considering as
recovered objects only those found within 0.5 pix of the given
coordinates, and with magnitudes differing from the input ones less than
$\pm$ 0.75 mag), and is given in Table~\ref{tab-compl}.  

The difference between input and output magnitudes of the artificial stars
provide a robust estimate of the photometric error to be
associated to each magnitude bin. These errors tend to be larger 
than the Daophot $\sigma$'s 
going towards fainter magnitudes. In Fig.~\ref{fig-ref1} we show the results
of our procedure for the three filters.  
These are also the errors adopted for the synthetic CMDs (see Section 4).

\begin{table}
\caption{Completeness function in the three bands; mag is intended as
$B$ or $V$, or $I$.}
\begin{tabular}{cccc}
\hline
Mag & compl. $B$ & compl. $V$ & compl. $I$ \\
\hline
12.75 &  -    &  -    & 1.000 \\
13.25 &  -    &  -    & 1.000 \\
13.75 &  -    & 1.000 & 1.000    \\
14.25 & 0.993 & 1.000 & 1.000 \\
14.75 & 0.991 & 1.000 & 1.000 \\
15.25 & 0.990 & 1.000 & 1.000 \\
15.75 & 0.989 & 1.000 & 1.000 \\
16.25 & 0.988 & 1.000 & 1.000 \\
16.75 & 0.989 & 1.000 & 1.000 \\
17.25 & 0.990 & 1.000 & 0.988 \\
17.75 & 0.992 & 0.997 & 0.985 \\
18.25 & 0.991 & 0.995 & 0.982 \\
18.75 & 0.990 & 0.997 & 0.978 \\
19.25 & 0.990 & 0.994 & 0.972 \\
19.75 & 0.989 & 0.992 & 0.963 \\
20.25 & 0.988 & 0.992 & 0.951 \\
20.75 & 0.986 & 0.994 & 0.936 \\
21.25 & 0.984 & 0.986 & 0.906 \\
21.75 & 0.980 & 0.985 & 0.810 \\
22.25 & 0.971 & 0.966 & 0.581 \\
22.75 & 0.936 & 0.916 & 0.289 \\
23.25 & 0.809 & 0.595 & 0.096 \\
23.75 & 0.540 & 0.082 & 0.023 \\
24.25 & 0.242 & 0.001 & 0.004 \\
24.75 & 0.072 &   -   &   -   \\
25.25 & 0.016 &   -   &   -   \\
\hline
\end{tabular}
\label{tab-compl}
\end{table}

\section{The colour - magnitude diagram}

The final, calibrated sample 
%{\it (tablefin\_bvi.radec.new)} 
contains 2239 stars; of them, 1195 have magnitudes in all three filters, 
35 have only $B$ and $V$  magnitudes, and 1009 have 
only $V$ and $I$ magnitudes. The $B$ exposures are the shallowest.
Pixel coordinates were transformed to $\alpha$ and $\delta$ using the
equatorial coordinates of 50 stars
all around the field, taken from the Digitized Sky Survey.\footnote{
The  DSS was produced at the Space Telescope Science Institute under U. S.
Government grant NAG W-2166.}
The table with the photometry and positions of all stars in our sample will
be made available through the BDA database
(Mermilliod 1995, {\em obswww.unige.ch/webda}).

We have made a star to star comparison of our magnitudes to PJM94's  (in the 
$V$ and $I$ bands), and Fig.~\ref{fig-conf} shows graphically the results: we
are brighter in $V$ by 0.063 mag, and fainter in $I$ by 0.119 mag.  
PJM94 do not give any details for their calibrations and the quality
of the nights when they observed (theirs is a survey work 
intended to search for interesting objects for dedicated follow up
observations, and was carried out also in marginally usable sky conditions).
Given also our good calibration, we feel rather confident of our figures. 

%%%%%%%%%%%%%%%%%%%%% Fig. 5 (CMDs: V,B-V ; V,V-I ; PJM94)
\begin{figure*}
\vspace{7cm}
\includegraphics{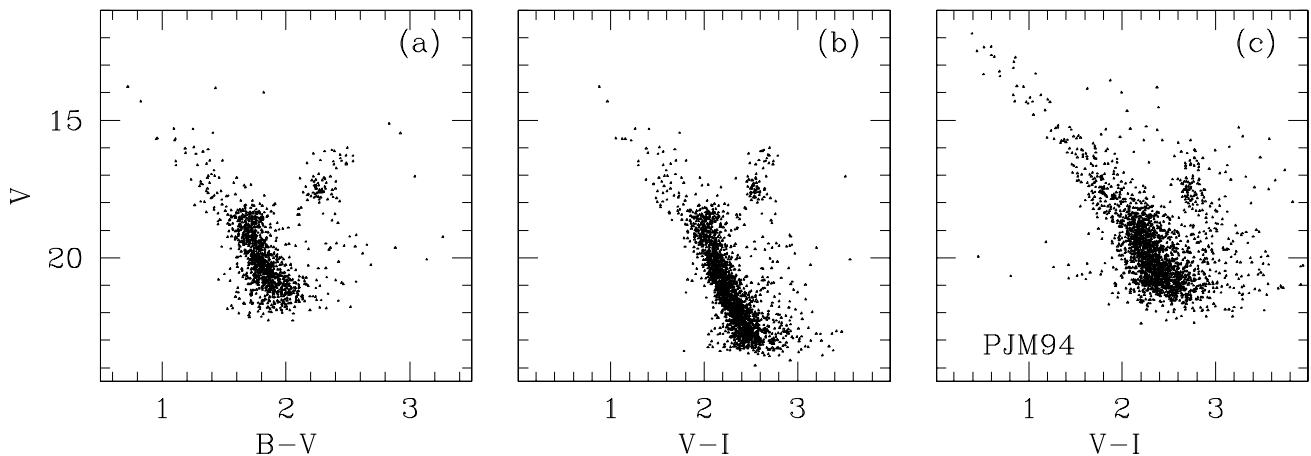}
\caption{Colour magnitude diagrams derived from our photometry:
(a) from the $B$ and $V$ frames, (b) from the $V$ and $I$ frames.
(c) Shown as a comparison, the only published diagram, by PJM94}
\label{fig-cmd}
\end{figure*}

%%%%%%%%%%%%%%%%%%%%%% Fig. 6 (external vs internal part PJM94)
\begin{figure}
\vspace{12cm}
\includegraphics{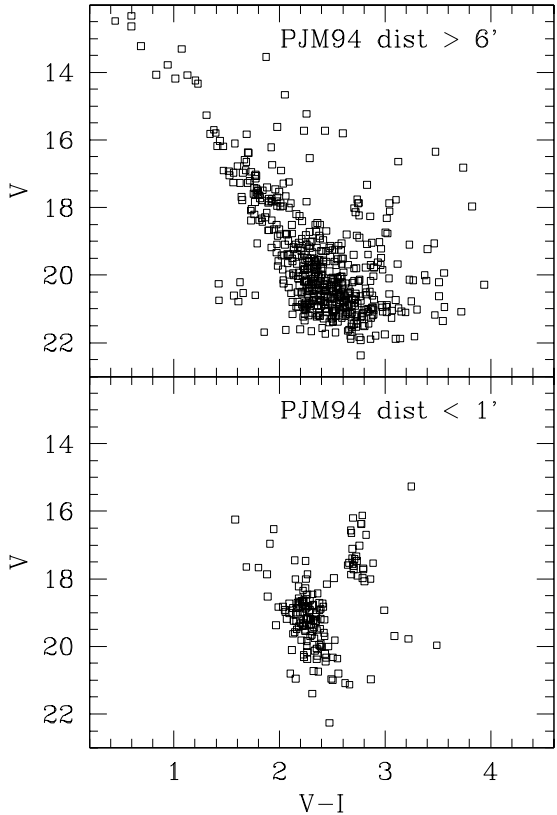}
\caption{Upper panel: CMD diagram taken from PJM94 for the outer part of the
cluster (distance greater than 6 arcmin from the centre). Lower panel: the same,
but for the innermost part (distance less than 1 arcmin)}
\label{fig-pjm3}
\end{figure}

%%%%%%%%%%%%%%%%%%%%%% Fig. 7 (map + histograms in X,Y)
\begin{figure}
\vspace{15cm}
\includegraphics{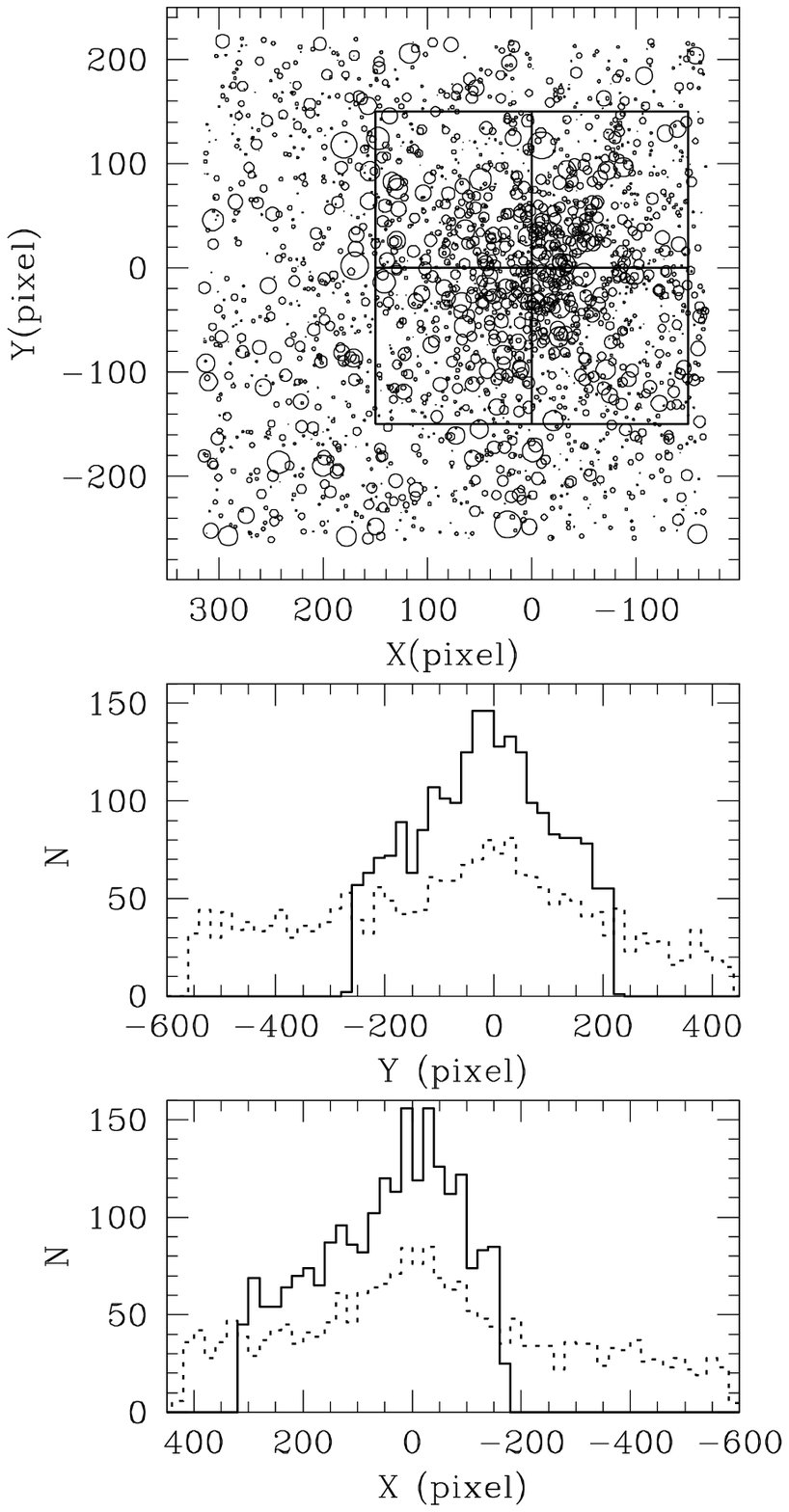}
\caption{Upper panel: Map of our field of view, based on photometry: each
pixel is 0.8 arcsec, the same scale of PJM94, that we adopted for
direct comparison.
The ``hole" around X $\simeq -80$, Y $\simeq 0$,
is caused by the blanketing we applied around a very bright star. Also
indicated are the four central quadrants (see text and Fig. 8).
Middle and lower panels: histograms of the Y and X positions of stars in
our field (solid line) and in PJM94's (dotted line): the centre position is
correct at the $\pm$ 10 pixel level.      
}
\label{fig-figref3}
\end{figure}

The CMDs resulting from our data are shown in Fig.~\ref{fig-cmd}(a) and (b),
while panel (c) shows for comparison PJM94's result (shallower, and based on a
larger field, hence more contaminated by field stars). Pismis 2 appears to be
an intermediate age cluster, heavily reddened but with both main-sequence,
MS, and red  clump (the locus of  central helium burning stars)
well populated and defined. The clump is characterized by
a fairly large extension in magnitude (see also Section 5). 
Since the structure of both the turn-off and the red clump  may be
complicated by differential reddening and binarity (see below), the
exact location in magnitude and colour of these features is not easy
to define.
For the clump, we place the position of the base, at its mean colour,
at $V=17.8$, $B-V$ = 2.22, and $V-I$ = 2.55. 
For the turn-off position we prefer to give a possible range: 
$V$ = 18.5 -- 18.8, $B-V$ = 1.5 -- 1.6, and $V-I$ = 1.8 -- 1.9.

We have no direct information on the cluster membership from proper 
motions or radial velocities. Hence, to separate ''cluster" and ''field"
objects in Pismis 2 we can only rely on the CMD appearance of an 
external control field, taken from PJM94 data, thanks to the BDA data base.

PJM94 observed the cluster with the CTIO 0.9m telescope, with a field of
view of 14 arcmin$^2$.
In the lower panel of Fig.~\ref{fig-pjm3} we show the CMD of their 
central 1 arcmin, while in the upper panel only stars farther than 6 arcmin 
from the centre are displayed:
it is immediately apparent that all the "blue" stars brighter than
V $\simeq$ 18 are most likely field objects, while all the red stars
brighter than V $\simeq$ 18 are probable cluster members.
To normalize PJM94's external field to our field of view, we have retained
of their 539 stars lying at distances larger than 6 arcmin only
the fraction corresponding to the ratio of our/their area. This provides
283 objects with the same overall distribution in the CMD as the 539 stars
shown the upper panel of Fig.~\ref{fig-pjm3}. 

There is a fairly large spread in the various sequences of the cluster CMDs of
Fig.~\ref{fig-cmd} (a) and (b).  
Two possible sources can be thought of: field star contamination and
differential reddening.

The cluster MS and the field stars MS appear to lie approximately at the same
colours in  Fig.~\ref{fig-cmd} (a) and (b),  but a small displacement of the
field star MS (e.g., due to the cluster being in front, hence  slightly bluer
and brighter, than the bulk of the field population)  could be present, and
explain the spread. However, the 283 field objects falling in our frame do not
seem to be sufficient to account for the whole spread in the sequences.
If we are witnessing the effect of field contamination, its importance  should
depend on the distance from the cluster centre, and we should see a
displacement between the MS at the centre and far from it. Unfortunately, the
study of the radial dependence of colour  is not really significative when
using our data, since the field of view is small and the cluster is not centred
on it. PJM94 data are taken on a larger field, but the photometric dispersion
in their CMD is so large that differences like those we are trying to see are
completely washed out. Anyway, if we look at the CMDs for stars  in the inner
arcmin and in circular shells 1 to 2, and 2 to 3 arcmin from the cluster
centre, the latter seem to present slightly bluer MS's, giving some support to
the hypothesis of field contamination as source of the spread.

\begin{figure}
\vspace{10cm}
\includegraphics{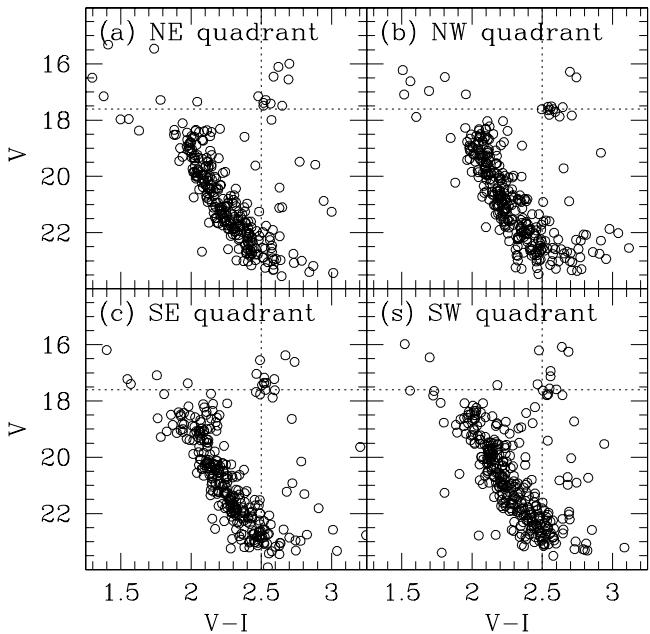}
\caption{
CMDs for the four central quadrants indicated in Fig. 7. The western ones are
slightly redder and fainter, as can be appreciated by the clump
position (the dashed line is shown here only to help the eye). The western
and eastern CMD can be overimposed by assuming a small difference in \ebv: 
$\Delta (B-V) = 0.04$ mag, i.e. $\Delta (V-I) = 0.05$, and 
$\Delta V =0.12$.  
}
\label{fig-figref4}
\end{figure}

An alternative explanation is the  presence of differential reddening in our
field of view, broadly speaking from the East to the West direction. To check
this possibility  we have plotted separately the CMDs of the four spatial
quadrants with origin in the cluster centre.  The map in the top panel of
Fig.~\ref{fig-figref3} shows the quadrant  locations; the histograms in the two
lower panels,  with the number of stars as a function of the X and Y
coordinates, indicate that the cluster centre is quite well determined (at the
$\pm$ 10 pixel level, i.e. 8 arcsec in the present scale). The top left CMD  of
Fig.~\ref{fig-figref4} refers to the stars in the North-East quadrant, the top 
right CMD to the NW quadrant, the bottom left CMD to the SE quadrant, and the
bottom right CMD to the SW quadrant. The number of stars in each region is
about 300 (the maximum difference is 10--15 per cent).  By comparing with each
other these four CMDs, one sees that the two East ones have very similar colour
and mag distributions, while the two West ones are similar to each other but
slightly redder than the East ones. Also the clump and MS turnoff luminosities 
appear slightly lower in the right than in the left panels. If one wants to
ascribe this difference to a reddening difference, the extra \ebv ~necessary to
bring both colours and luminosities into agreement is  $\Delta E(B-V) =
0.04\pm0.02$, i.e. less than 3 per cent of the reddening attributed to the
cluster by PJM94. In this case, the spread in the total cluster CMD would be
caused by the superpositions of two regions (the eastern  and the western ones)
affected by a different reddening. This possibility is supported by the
circumstance that larger scale maps of the sky region  definitely show a lower
number of stars in the area on the West side of Pismis 2, as if an additional
obscuring cloud was preventing the detection of faint objects (we inspected a
30$\times$30 arcmin$^2$ region of the DSS, but the effect is also visible in
our Fig.~\ref{fig-map}). If the obscuring  cloud is closer than Pismis 2 and
its outskirts cover also part of the  cluster (its western half), then the
affected stars appear redder and  fainter than the unaffected ones.

The existence of this ''hole'' in the map, combined with the fact that
not only the MS but also the clump (where no field stars are expected) present
colour and magnitude differences consistent with a differential reddening
cause, makes us prefer this as the most likely explanation for the
observed spread. Besides, the relative variation is so small (3 per cent
of the reddening) that it looks almost natural to have it.

%%%%%%%%%%%%%%%%% Fig. 8 (CMDs in the four central fields)

%%%%%%%%%%%%%%%%%% Fig. 9 (simulation single reddening)
\begin{figure}
\vspace{12cm}
\includegraphics{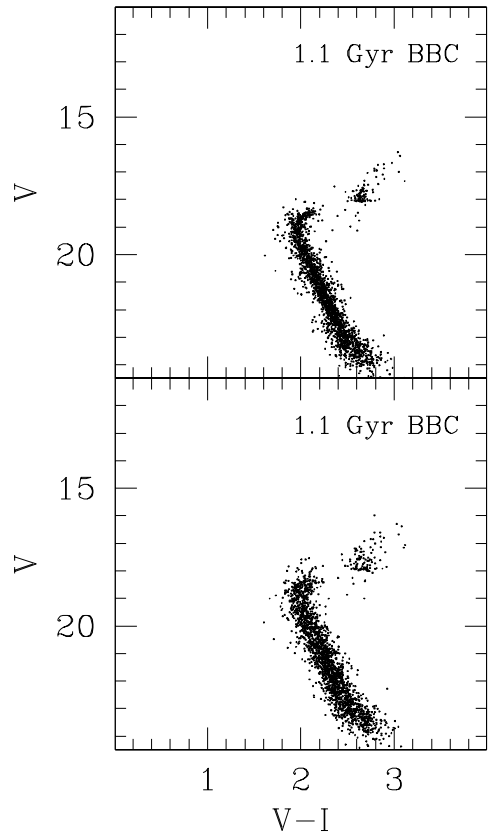}
\caption{Synthetic CMD obtained with the BBC models with solar metallicity,
age 1.1 Gyr, $(m-M)_{0}$=12.7. The top panel shows the result of assuming
a single reddening $E(B-V)$=1.29, the bottom panel that of assuming that
50 per cent
of the stars have $E(B-V)$=1.26 and the other 50 per cent have $E(B-V)$=1.32. In
the bottom panel we also assume that half of the stars are binaries.
}
\label{simpap}
\end{figure}

\section{Cluster parameters}

In order to derive the age, distance and reddening for Pismis 2, we have
applied our usual method of comparison of the observed CMD to synthetic
CMDs based on homogeneous sets of stellar evolution models. As described
in Tosi et al. (1991) and in the previous papers of this series, the method
consists in creating, via Monte Carlo extractions on the stellar tracks,
synthetic CMDs which must contain the same number of stars as the observed
diagram and must suffer of the same photometric errors and incompleteness
factors as derived from the real data. 
We have adopted as reference CMD the diagram shown in Fig.~\ref{fig-cmd} (b), 
which contains the 2204 objects measured and cross-identified in $V$ and $I$. 
To take into account the background/foreground contamination estimated in 
the previous Section, the synthetic CMDs for Pismis 2 contain 1920 stars. 
Since the $B$ frames were not as deep as the $V$ and $I$ ones, the $V, B-V$ 
CMD [Fig.~\ref{fig-cmd} (a)], is shallower and we have therefore used it only 
as further check of the synthetic models. 

The extracted synthetic stars have luminosity and temperature given by the
adopted set of stellar models, and are then placed into the observational 
colour--magnitude plane using the Bessel, Castelli \& Pletz (1998) photometric 
conversion tables. They are attributed a photometric error derived
from the magnitude distribution of (input--output) mags of the artificial
star tests performed to derive the incompleteness factors of our photometry
(Section 2.2). They are retained or rejected according to a completeness
test based on the empirical incompleteness factors. 
The magnitudes and colours of the 1920 survived synthetic stars
are modified according to the assumed reddening and distance modulus.
Since our photometry is calibrated into the Johnson-Cousins system, we
assume $E(V-I)=1.25 \times E(B-V)$ (Dean, Warren \& Cousins 1978).

To test the intrinsic uncertainty related to the stellar models, we perform 
the CMD simulations adopting evolutionary sets from different groups.
We have used the Padova (e.g. Bressan et al. 1993, hereinafter BBC), 
FRANEC (Dominguez et al. 1999, hereinafter FRA), and FST (Ventura,
D'Antona \& Mazzitelli 2001, in preparation) models of solar and sub-solar
metallicities. Among the FST models we have adopted those with overshooting
parameter $\eta$=0.2. For Pismis 2, the results from the different sets 
are strikingly consistent with each other. We
find that, independently of the adopted set, only the models with solar 
metallicity provide MS slopes in agreement with the data.

A common feature of all the synthetic diagrams is that, despite
the fairly large size of the adopted photometric errors, their stellar
distributions are always tighter that the empirical one. 
This effect is shown in Fig.~\ref{simpap}, where the top panel
shows the synthetic CMD corresponding to the BBC solar models with the
best fitting parameters: age 1.1 Gyr, $(m-M)_{0}$=12.7 and $E(B-V)$=1.29.
Even assuming that a large fraction of the cluster members are binary stars
(with random mass ratio), we cannot reproduce the observed spread. 
Given the discussion on the spread of the evolutionary sequences in Section 3,
we decided to adopt two and not just one reddening.
To get the required distribution (bottom 
panel of Fig.~\ref{simpap}), we need to assume that 50 per cent of the stars 
are affected by a reddening $\Delta(E(B-V))$=0.06 higher than that affecting 
the other 50 per cent. 
This implies a fractional variation of only 5 per cent, which is
not implausible in a region so highly extincted. 

Independently of the adopted set of stellar models, all the synthetic CMDs in 
better agreement with the reference diagram turned out to be those with solar 
metallicity and assuming $E(B-V)$=1.26 for 960 objects and $E(B-V)$=1.32 for 
the remaining 960 objects. They also assume that each of the two populations 
contains a 50 per cent of binary stars with equally distributed mass ratios. 

These reddening values are lower than the $E(B-V)$=1.48 derived by PJM94.
The difference can be totally ascribed to the different calibration, since
our $V-I$ colours have turned out (Section 3) to be 0.18 mag bluer than theirs.
Nonetheless, we have simulated several synthetic diagrams with sub-solar
metallicity to check whether a lower metal content could simultaneously
provide a fair agreement with our data and with PJM94's reddening. We have
found, instead, that at $Z\simeq0.5Z_{\odot}$ the synthetic MS slopes
become inconsistently flatter than the observed one, while the required 
reddening is still lower than PJM94's.\footnote{For instance, with BBC at 
$Z=0.008$ the two necessary reddenings are $E(B-V)=1.34-1.40$, with FRA at 
$Z=0.01$, they are $E(B-V)=1.33-1.39$.}

The CMDs of the cases in better agreement with the reference diagram for 
each set of solar metallicity models are shown in the top panels of 
Fig.~\ref{simpap2}, while the corresponding luminosity functions (LFs) are 
compared with the empirical one in the bottom panels. Since the reference
CMD of Fig.~\ref{fig-cmd} (b) includes also non-member stars,   we have added
to the  synthetic stars the 283 objects contained in the normalized external
field  described in Section 3, for a better comparison between synthetic and
empirical data.
The external field, derived from PJM94, is shallower
than ours, but still provides a good representation of the actual 
contamination.
Also the cluster LF (dots) is derived from the whole reference CMD and we
have therefore considered the 283 external stars also in the derivation
of synthetic LFs (curves).

We obtain an age of 1.1 Gyr with the BBC models, and of 1.2 Gyr with the
FST ones. The FRA models provide the best curvature of the MS, but don't
reproduce equally well the turn-off region, which makes it more difficult
to distinguish the best age. Ages between 1.0 and 1.2 Gyr lead to 
comparable agreement with the data, thus we show in the figure the
CMD corresponding to the mean age, 1.1 Gyr.
We obtain a distance modulus of $(m-M)_{0}$=12.7 with
BBC and FRA, and $(m-M)_{0}$=12.5 with FST. Considering also the coincident
reddening values obtained with the three sets, we have a striking and
fairly unusual agreement in the derived values of all the parameters.

The synthetic CMDs in better agreement with the data reproduce fairly well 
the observed features of both MS and evolved stars. Also the group of red
stars brighter than the clump is naturally predicted, since the total number
of observed cluster members if large enough to let also the faster core He -
burning phases to be clearly represented in the CMDs. 
Pismis 2 falls in the age range where Girardi, Mermilliod \& Carraro
(2000) found necessary  to invoke for some clusters an additional mechanism,
like differential mass loss on the RGB, or rotation, to explain the
clump configuration. However, our data do not require additional mechanisms
to be reproduced by synthetic CMDs.

Finally, we have cross-checked our results by comparing the $V, B-V$ CMD
of all the synthetic cases with the diagram of  Fig.~\ref{fig-cmd} (a).
This diagram contains the 1230 objects for which $B$ and $V$ were measured and
is much more incomplete than the reference $V, V-I$ CMD, because
the $B$ frames are significantly shallower. Hence, the models evaluation in
this case is restricted to the morphological distribution in the CMD, and
cannot take into account the number of stars in the various evolutionary
phases. We find that all the models in better agreement with the $V, V-I$
CMD are in agreement also with the $V, B-V$ distribution, but only if we
make the $B$ fluxes 0.07 mag dimmer. This does not mean that we invoke
different reddening or distance modulus, but that our $B$ mags are not
completely consistent with the corresponding $V$ and $I$ mags for what concerns
the stellar physical properties. We ascribe the $\Delta B$=0.07 necessary
to bring the data back to self-consistency to calibration difficulties in
the $B$ band. As discussed in Section 2.1, to obtain a calibration as safe
and accurate as possible, we have considered for the $V$ and $I$ bands not
only the usual Landolt's standards but also Stetson's extension to fainter
objects. This has allowed us to cover a wider range of the observed colours
with the standards used for the calibration. This extension has not
been possible for the $B$ band, where the available standards cover only
half of the observed colour range. Hence, its calibration is inevitably 
more uncertain than that in the other two bands and is the most likely
cause of the $\Delta B$=0.07 offset. Fig.~\ref{simpap3} shows the synthetic
$V, B-V$ diagram corresponding to the same BBC case as the $V, V-I$ of
the top left panel of Fig.~\ref{simpap2}, already corrected for $\Delta B$=0.07.

\section{Summary and conclusions}

%%%%%%%%%%%%%%%%%%%%% Fig. 10 (simulations, best fit)
\begin{figure*}
\vspace{12cm}
\includegraphics{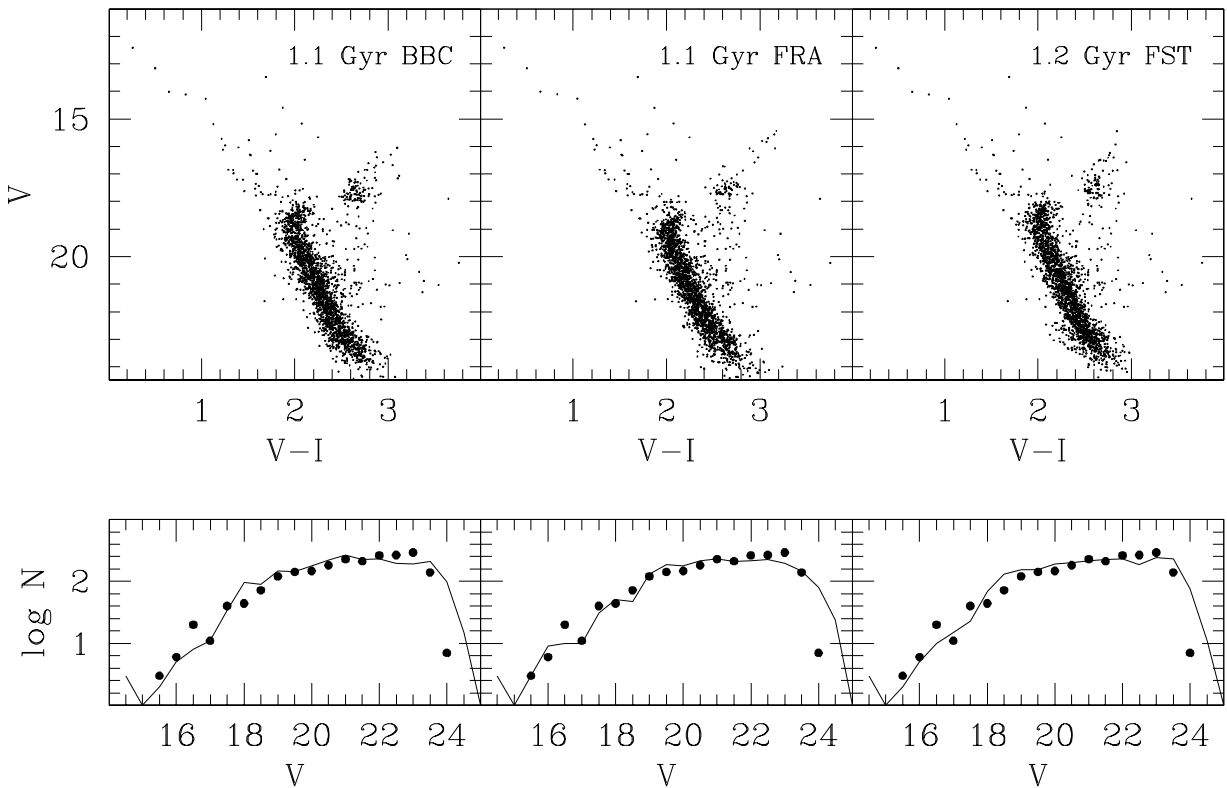}
\caption{Best models resulting from the BBC (left panels), FRA (central panels)
and FST (right panels) tracks. All of them have solar metallicity, and assume
$E(B-V)$=1.26 and $E(B-V)$=1.32. The distance modulus is $(m-M)_{0}$=12.7
for BBC and FST, and 12.5 for FRA. The age is indicated in the top panels.
The corresponding global LFs (curves) are compared with the empirical
one (dots) in the bottom panels.
}
\label{simpap2}
\end{figure*}

We have presented the first detailed study of Pismis 2, an intermediate age
open cluster, located towards the external part of our Galaxy.
Pismis 2 is seated in a highly reddened Galactic 
region, probably inside a dust layer, as discussed by Dutra \& Bica (2000).
These authors attribute an excessive reddening to the cluster, derived by
extrapolating into the forbidden region the absorption maps by Schlegel et al.
(1998) and place the layer at 3.8 kpc from the sun. We find smaller reddenings
and shorter distance (about 3.3 kpc), but we agree on the 
overall conclusion of a heavily
obscured zone. Besides, we suggest that the absorbing clouds are not evenly
distributed on the sky area around the cluster, because we find some evidence
for a $\Delta E(B-V)$=0.04--0.06 across the cluster field of view.  

The high and possibly
differential reddening affecting Pismis 2 prevents its CMD
from clearly showing the exact shape and location of the MS turn-off or
some of the typical features of other open clusters, like MS gaps, binary
stars sequence, etc. Nonetheless, the parameters derived from the simulated
CMDs and based on stellar models from various groups are in good agreement
with each other, thus suggesting that, despite the fairly large spread of
the observed CMD, the uncertainty on their values is not larger than for
other clusters. In fact, even if we cannot distinguish morphological
details, the synthetic CMDs are quite well constrained by the number and
the mag-distribution of the stars in the clump, by the mag-difference 
between the clump and the MS top, by the MS slope.
We thus derive an age of 1.1 -- 1.2 Gyr (confirming PJM94 result) using three
different sets of theoretical tracks. We do not find any compelling reason
to prefer one set over the others, and with all of them the best fits are
obtained assuming a 50 percent of binary systems. The  true distance
modulus is found to be 12.5 -- 12.7. The metallicity is probably about 
solar, but more direct measurements via spectroscopy or narrow band 
photometry are necessary to confirm it.

The derived age is similar, although slightly older, than that (0.75--1.0
Gyr, depending on the adopted models, see Sandrelli et al. 1999) 
of NGC 2660, one of the clusters already examined by our group.
Indeed, the CMDs of the two clusters have quite
similar features. NGC 2660 has a much tighter distribution (see fig. 4
of Sandrelli et al.), thanks to the lower and single 
reddening $E(B-V)\simeq$0.4, but the slope and the shape of the MS,
as well as the morphology of the turn-off region, resemble very closely
those of Pismis 2. Also the relative colour difference between the MS
turnoff and the clump are the same in the two clusters, both in $B-V$ and
in $V-I$. What changes from one another is the number and the magnitude
distribution of the clump stars: Pismis 2 has a much more populated clump
and a slightly larger $\Delta V$ between the MS turnoff and the clump basis.
$\Delta V$ is a well known age indicator (MAI, see PJM94) and the higher value
confirms that Pismis 2 is slightly older than NGC 2660. 

The clump of Pismis 2 spans two magnitudes, while the one of NGC 2660
is populated only in its fainter portion, but both can  be
reproduced by the natural range of masses present in a single isochrone.
Since the faint edge of the clump 
is the evolutionary phase where core-He-burning stars stay longer, part of the
difference could be simply due to the much larger number of stars resolved in
Pismis 2 (2206) than in NGC 2660 (407).
The $V$ LFs of the two clusters, when normalized to the respective number of
resolved stars, are so similar to each other that they completely overlap
at all magnitudes, except at the base of the clump and at the faintest bins.
The former difference is the signature of the $\Delta V$ -- age relation, and
the latter of the different incompleteness levels of the two data sets.

At odds with what was found for many other open clusters (e.g., in NGC
6253, Bragaglia et al. 1997, and examples cited there), we do not observe any
noticeable flattening of the lower main sequence luminosity function. In
Fig.~\ref{simpap2}, lower panels, the filled symbols representing the empirical
data seem to decrease at $V \simeq 23$ only because of the sharp change in the
completeness of our measurements (see Table~\ref{tab-compl}). Pismis 2
does not appear to have suffered much from selective evaporation of low
mass stars.

\bigskip\noindent

ACKNOWLEDGEMENTS

We warmly thank P. Montegriffo, whose programs were used for the data analysis.
The bulk of the simulation code was originally provided by L.Greggio.
We thank F. D'Antona and P. Ventura for providing their unpublished
stellar models.
Finally, we wish to thank the referee, J.-C. Mermilliod, for useful comments
that significantly improved the paper.
This research has made use of the Simbad database, operated at CDS,
Strasbourg, France. 
We also acknowledge the use of the valuable BDA database, maintained by J.-C.
Mermilliod, Geneva.
L.DF. thanks the Bologna Observatory for the financial support.

%%%%%%%%%%%%%%%%%%%% Fig. 11 (simulation, B-V)
\begin{figure}
\vspace{7cm}
\includegraphics{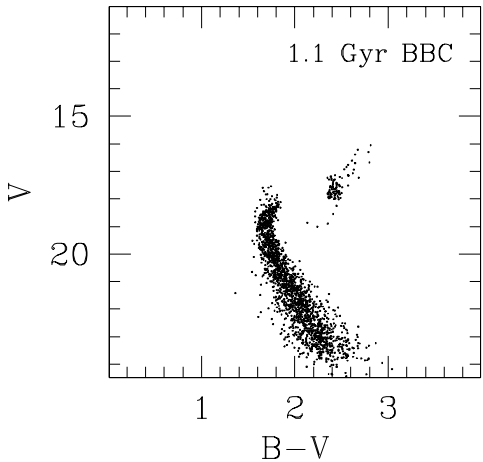}
\caption{Synthetic $V, B-V$ diagram corresponding to the best BBC case
of Fig.~\ref{simpap2}.}
\label{simpap3}
\end{figure}


\begin{thebibliography}{}

\bibitem{} Bessel, M.S., Castelli, F., Plez, B. 1998, A\&A, 337, 321

\bibitem{} Bragaglia, A., {\it et al.} 2001, AJ, 121, 327

\bibitem{} Bragaglia, A., Tessicini, G., Tosi, M., Marconi, G., Munari, U.
 1997, MNRAS, 284, 477
 
\bibitem{} Bragaglia, A., Tosi, M., Marconi, G., Carretta, E. 2000,
 F. Matteucci, F. Giovannelli, eds,
 The chemical evolution of the Milky Way: Stars versus clusters,  
 Kluwer Academic Publishers (Dordrecht), 255, p. 281

\bibitem{} Bressan, A., Fagotto, F., Bertelli, G., Chiosi, C. 1993,
 A\&AS, 100, 647

\bibitem{} Carretta, E., Bragaglia, A., Tosi, M., Marconi, G. 2000, 
 R. Pallavicini, G. Micela, S. Sciortino, eds,
 Stellar clusters and associations: convection, rotation, and dynamos,
 ASP Conf.Ser. 198, p. 273

\bibitem{} Dean, J.F., Warren, P.R., Cousins, A.W.J. 1978, MNRAS, 183, 569

\bibitem{} Dominguez, I., Chieffi, A., Limongi, M., Straniero, O. 1999,
 ApJ, 525, 226

\bibitem{} Dutra, C.M., Bica, E. 2000, A\&A, 359, 347

\bibitem{} Friel, E.D. 1995, ARAA, 33, 38

\bibitem{} Girardi, L., Mermilliod, J.-C., Carraro, G. 2000, A\&A, 354, 892

\bibitem{} Janes, K.A., Phelps, R.L. 1994, AJ, 108, 1773

\bibitem{} Landolt, A.U. 1992, AJ, 104, 340

\bibitem{} Mayor, M. 1976, A\&A, 48, 301

\bibitem{} Mermilliod, J.C. 1995, D. Egret, M.A. Albrecht, eds, Information and
 On-Line Data in Astronomy, Kluwer Academic Press (Dordrecht), p. 127

\bibitem{} Miller, N.A., Hing, L.N., Friel, E.D., Janes, K.A. 1995, 
 AAS, 187, 10704

\bibitem{} Panagia, N.,  Tosi, M. 1981, A\&A, 96, 306

\bibitem{} Patino, R.M., Friel, E.D. 1994, AAS, 185, 10306

\bibitem{} Phelps, R.L., Janes, K.A., Montgomery, K.A. 1994, AJ, 107, 1079
 (PJM94)

\bibitem{} Pismis, P. 1959, BOTT, 2, 37 
 %(Bol.Obs.Tonantz.Tacub.,2,part no 18,37)

\bibitem{} Sandrelli, S., Bragaglia, A., Tosi, M.,  Marconi, G. 1999,
 MNRAS, 309, 739

\bibitem{} Schlegel, D.J., Finkbeiner, D.P., Davis, M. 1998, ApJ, 500, 525

\bibitem{} Stetson, P.B. 1992, User's Manual for DAOPHOT-II

\bibitem{} Stetson, P.B. 2000, PASP, 112, 925

\bibitem{} Tosi, M., Greggio, L., Marconi, G., Focardi, P. 1991, AJ, 102, 951

\end{thebibliography}
\end{document}